\documentstyle[aps,multicol,epsf]{revtex}

\newcommand{\be}{\begin{equation}}
\newcommand{\ee}{\end{equation}}

\begin{document}
%\preprint{SNUTP 98-132}
\draft
\widetext

\title{Intrinsic finite-size effects in the two-dimensional $XY$ model\\
with irrational frustration}

\author {Sung Yong Park$^{1,2}$, M.Y. Choi$^2$, Beom Jun Kim$^{3}$,
Gun Sang Jeon$^4$, 
and Jean S. Chung$^{5}$} 

\address {$^1$Lyman Laboratory of Physics, Harvard University, Cambridge, 
Massachusetts 02138\\
$^2$Department of Physics, Seoul National University, Seoul 151-742, Korea\\
$^3$Department of Theoretical Physics, Ume{\aa} University, 901 87 Ume{\aa}, Sweden\\
$^4$Center for Strongly Correlated Materials Research, Seoul National University, Seoul 151-742, Korea\\
$^5$Department of Physics, Chungbuk National University, Cheongju 361-763, Korea}

\maketitle
\begin{abstract}
This study investigates in detail the finite-size scaling of the 
two-dimensional irrationally frustrated $XY$ model. 
By means of Monte Carlo simulations with entropic sampling, we examine the 
size dependence of the specific heat, and find remarkable deviation from the 
conventional finite-size scaling theory, which reveals novel intrinsic 
finite-size effects.
Relaxation dynamics of the system is also considered, and correspondingly,
finite-size scaling of the relaxation time is examined, 
again giving evidence for the intrinsic finite-size effects and suggesting
a zero-temperature glass transition.
\end{abstract}


\begin{multicols}{2}
\narrowtext
The two-dimensional $XY$ model with uniform frustration has attracted much 
interest in connection with quite a few physical realizations such as
periodic arrays of coupled Josephson junctions in uniform magnetic 
fields~\cite{review}.
Of particular interest is the case of irrational frustration, where
the possibility of a structural glass phase without intrinsic random disorder 
has been a long-standing question.
In the model with irrational frustration, 
the periodicity of the ground state should be incommensurate with the 
underlying array periodicity and the ground states are infinitely degenerate,
suggestive of the absence of a finite-temperature transition.
It was thus argued that there is no phase transition at finite temperatures in
the thermodynamic limit~\cite{MYChoi85}, which was also favored by numerical 
simulations of the current-voltage ($IV$) characteristics~\cite{Granato}.  
The latter, performed with resistively shunted junction (RSJ) dynamics,
demonstrated that the $IV$ scaling is consistent with a zero-temperature 
vortex-glass transition proposed in Ref.~\cite{DSFisher91}.
In other simulations, on the other hand, a metastable 
finite-temperature glass transition was proposed~\cite{Halsey,quasi} 
and the similarity of the relaxation dynamics to the 
primary relaxation in a supercooled liquid was pointed out~\cite{BKim}.
Subsequent Monte Carlo (MC) simulations of the original $XY$ 
model~\cite{Denniston} and of the 
corresponding Coulomb gas~\cite{Gupta} then suggested 
the relevance of boundary conditions and dynamics in observing
the reported glass effects, leaving the existence of 
a finite-temperature transition still inconclusive.

One way to resolve this is to attribute the observation of the finite transition
temperature to {\em intrinsic} finite-size effects.
In a finite system the effects of irrational frustration 
may not be distinguished from those of rational frustration, 
the value of which is given by a rational approximant sufficiently 
close to the irrational~\cite{MYChoi87}.
Consequently, unlike the system in the thermodynamic limit,
a finite system (with irrational frustration) tends to behave as if the value of 
the frustration were given by the rational approximant.
Accordingly, the finite system may display such transition-like behavior 
as the specific heat peak (rounded off due to usual finite-size effects) 
at a finite temperature, 
even though the system in the thermodynamic limit undergoes a zero-temperature 
transition, as argued in Ref.~\cite{MYChoi85}.
Here the approximant which properly describes a finite system 
depends on the system size~\cite{MYChoi87}, raising
the interesting possibility of intrinsic finite-size effects
in addition to the usual ones;
this has not been probed in previous simulations, 
where, for simplicity, fixed
rational approximants were mostly used instead of the actual
irrational value of frustration. 

In this Letter, we adopt the idea of the intrinsic
finite-size effects to resolve the controversy on the phase transition in 
the two-dimensional irrationally frustrated $XY$ (IF$XY$) model. 
For this purpose, we use the exact irrational number
up to the machine precision, and perform extensive MC simulations with
improved entropic sampling~\cite{es} to compute the specific heat.
The obtained data turn out to be inconsistent with the conventional 
scaling theory of 
either the first-order transition or the second-order one, 
revealing the presence of novel intrinsic finite-size effects. 
To confirm the existence of such intrinsic 
finite-size effects, we also study the relaxation dynamics of the system 
and construct the finite-size scaling theory of the relaxation time,
on the basis of the zero-temperature glass transition 
together with the intrinsic finite-size effects.
The size dependence of the relaxation time indeed agrees well with the
scaling theory, providing strong evidence for the intrinsic finite size effects.

The Hamiltonian for the irrationally frustrated $XY$ model on an $L\times L$ 
square lattice is given by
\begin{equation}
H=-J\sum_{\langle i,j\rangle}\cos(\phi_i-\phi_j-A_{ij}),
\end{equation}
where the angle $\phi_i$ corresponds to the phase of 
the superconducting order parameter
at site $i$ and $J$ represents the Josephson coupling strength in the case of
a Josephson-junction array.  The plaquette sum of the bond angle
$A_{ij}$ is constant over the whole lattice, $\sum A_{ij} = 2\pi f$, where
the frustration parameter $f$ is set equal to the golden mean: 
$f=\Omega \equiv (\sqrt{5}-1)/2$. 

To investigate the finite-size effects of this system, we consider systems
of size $L= 5, 8, 13, 21,$ and $34$, the entropies of which
are computed by means of extensive MC simulations with entropic sampling.
The entropy obtained for each system allows one to calculate the specific heat of the system.
Both periodic boundary conditions (PBC) and free boundary conditions (FBC)
are used and the results are compared with each other, from which two
types of finite-size effects are identified.
In principle both boundary conditions should yield correct limiting behavior 
as the system size $L$ is increased,
although for finite $L$ the PBC induces mismatch at the boundary.
The values of $L$ are thus chosen in such a way that the mismatch
present in the PBC is kept minimum. 
%%%%%
In the case of the golden mean, it
can be achieved by choosing the Fibonacci numbers. 
%%%%%

\begin{figure}
\centerline{\epsfxsize=8cm \epsfysize=6cm  \epsfbox{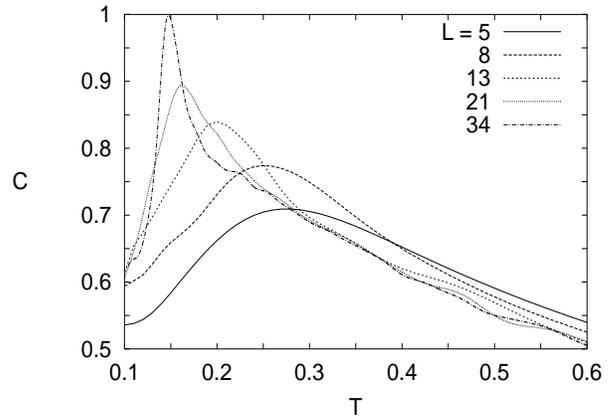}}
\caption[Specific heat versus temperature in the irrationally frustrated XY model]
{Specific heat versus temperature in the irrationally frustrated XY model
under periodic boundary conditions, for size $L=$ 5, 8, 13, 21, and 34. The peak 
position shifts toward lower temperatures as the system size is 
increased.}
\label{fig:IfXY specific heat}
\end{figure}

Figure~\ref{fig:IfXY specific heat} presents the temperature dependence of the
obtained specific heat in the PBC for various system sizes.
For $L=34$, in order to get reliable results, we have performed as many as 
$7.5\times 10^{10}$ MC steps per site using a Cray T3E supercomputer system.
As $L$ gets increased, the peak is shown to grow and the transition region 
to narrow whereas the peak position shifts toward lower temperatures.
The position of the specific heat peak in the system of size $L$, 
denoted by $T_m(L)$, is expected to approach the critical 
temperature $T_c$ as $L$ is increased.
The well-known finite-size scaling theory of the second-order 
transition~\cite{Cardy,Privman} gives the behavior
\begin{equation}
T_m(L)-T_c\propto L^{-1/\nu}, \label{eq:so_finite_size_effect}
\end{equation}
to which the least-square fit of our data
leads to $T_c=0$ and $1/\nu=0.36$.
%%%%%
On the other hand, the peak height in Fig.~\ref{fig:IfXY specific heat}, 
expected to grow as $L^{\alpha/\nu}$, gives the value $\alpha/\nu \approx 0.17$,
which, via the Josephson scaling relation 
($2-\alpha =d\nu$ with the system dimension $d=2$), leads to the value 
$\nu \approx 0.92$; this is consistent with 
the value $\nu \approx 0.9$
obtained in the numerical study of $IV$ characteristics~\cite{Granato}.
%%%%%
Thus the result of the fit manifests inapplicability 
of the conventional finite-size scaling theory.
Further, it is also inconsistent with the scaling theory of the first-order 
transition~\cite{Privman,Peczak}, according to which
%\begin{equation}
%T_m(L)-T_c\propto L^{-d} \label{eq:first_order_FSS1}
%\end{equation}
$1/\nu$ in Eq.~(\ref{eq:so_finite_size_effect}) is replaced by
the system dimension $d=2$. 
It is thus concluded that the system does not display conventional 
finite-size effects associated with either a first-order transition 
or a second-order one.
Still remains the possibility of a weakly first-order transition,
in which case the system may appear to exhibit size dependence similar to that 
for a second-order transition~\cite{Privman,Peczak}. 
Such pseudo-divergent scaling behavior should disappear
as the system size exceeds the correlation length of the system.

\begin{figure}
\centerline{\epsfxsize=8cm \epsfbox{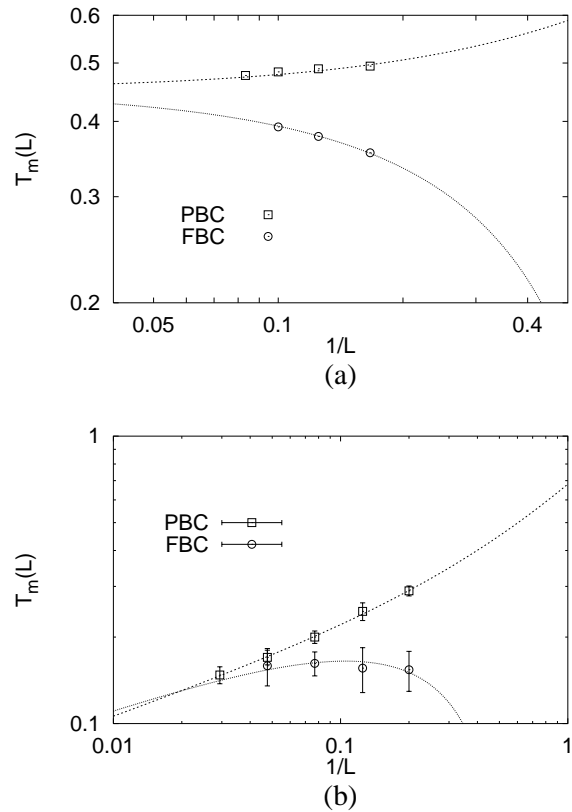}}
\caption
{Size dependence of the position of the specific heat peak. (a) 
In the rational case ($f=1/2$), the data in both PBC and FBC follow 
Eq.~(\ref{eq:so_finite_size_effect}), with the lines denoting the least square 
fits to Eq.~(\ref{eq:so_finite_size_effect}). 
(b) In the irrational case ($f=\Omega$), the data fit to 
Eq.~(\ref{eq:PBC scaling}) (PBC) or (\ref{eq:FBC scaling}) (FBC) rather than to
Eq.~(\ref{eq:so_finite_size_effect}).  In (a) error bars are not larger than the symbols.}
\label{fig:peak}
\end{figure}

%%%%%
Deviation from these conventional 
(second-order, first-order, or weakly first-order)
transitions can be observed most clearly
in comparing the size dependences of the specific heat 
under PBC and under FBC.
Periodic boundary conditions tend to suppress fluctuations, 
exhibiting the peak position
at a temperature higher than the true critical temperature $T_c$.
Under FBC, on the other hand, no such suppression exists and large fluctuations
in general yield the peak at a lower temperature~\cite{Cardy}.
Thus the ordinary finite-size effects are manifested by the
shift of the peak position toward $T_c$ as the system size is increased, 
from higher temperatures in PBC and from lower temperatures in FBC, respectively.
The results of test runs for the fully frustrated $XY$ model with $f=1/2$, 
shown in Fig.~\ref{fig:peak}(a), 
indeed demonstrate such behavior of the ordinary finite-size effects: 
The data in both cases, PBC and FBC, follow
Eq.~(\ref{eq:so_finite_size_effect}) (up to the overall sign),
and the least-square fit with the values of $T_c$ and $\nu$ in Ref.~\cite{Olsson}
yields the lines displaying nice scaling behaviors~\cite{footnote1}.
On the other hand,
Fig.~\ref{fig:peak}(b) reveals quite different features 
of the finite-size effects in the IF$XY$ model.
Here the peak position $T_m(L)$ in PBC is sensitive to the system size $L$,
reducing rapidly with $L$,
while it is not the case in FBC~\cite{footnote}.
In particular, unlike the case of rational frustration, 
%where the peak positions of both boundary conditions 
%approach a middle value between two values,
the peak position in FBC does not shift toward higher temperatures, 
violating the conventional scaling given by Eq.~(\ref{eq:so_finite_size_effect}).

Such peculiar behavior can be explained in terms of intrinsic finite-size effects
present in addition to the ordinary ones~\cite{MYChoi87}: 
For the frustration given by the golden mean, the appropriate rational
approximants are given by the ratios between adjacent pairs in the Fibonacci
sequence, i.e., $f_k=q_{k-1}/q_k$, where $q_k$ denotes the $k$th Fibonacci number.
Among these rational frustration values $f_k$'s, 
those satisfying the following condition
determine nature of the transition in the (finite) system of size $L$:
\begin{equation}
q_k\ll L\ll \left|f-f_k\right|^{-1},
\label{eq:size_criterion}
\end{equation}
where $f=\Omega$.  While the first inequality in 
Eq.~(\ref{eq:size_criterion}) simply implies that the system size should be
sufficiently larger than the size of a unit cell, the second one,
related with the charge neutrality condition, gives the
criterion that the defects due to the mismatch $f-f_k$ are negligible since 
the corresponding phase change across the system is 
$2\pi L (f-f_k)$~\cite{MYChoi87}.
Accordingly, as the system size $L$ is increased, 
the mismatch should get smaller and the bigger Fibonacci number $q_k$ 
is needed to describe the transition at the size.
The transition temperature as well as nature of the transition 
depends on the frustration parameter or on $q_k$~\cite{review,MYChoi85,choi85}
and thus on the system size. 

%%%%%
It is therefore plausible to replace $T_c$ in Eq.~(\ref{eq:so_finite_size_effect}) by
$T_c (L)$, the transition temperature of the system of size $L$,
which yields 
\begin{eqnarray} \label{eq:bc_finite_size_effect}
T_m(L)-T_c(L) &=& c_p L^{-1/\nu} \nonumber \\
T_m(L)-T_c(L) &=& -c_f L^{-1/\nu} 
\end{eqnarray}
with appropriate (positive) amplitudes $c_p$ and $c_f$ 
for PBC and for FBC, respectively.
Here $T_c(L)$ is expected to decrease with the system size, approaching the
true critical value $T_c$ in the thermodynamic limit~\cite{MYChoi87}. 
Since the precise limiting behavior is unknown, 
we assume the simple behavior
\begin{equation}
T_c(L)-T_c = c_i L^{-a}. \label{eq:tc_finite_size_effect}
\end{equation}
Note that in contrast to the ordinary finite-size effects, 
which depend crucially upon the boundary conditions,
the intrinsic effects described by Eq.~(\ref{eq:tc_finite_size_effect}) 
change the transition temperature itself, 
regardless of the boundary conditions.
%Namely, the intrinsic effects are expected to lower the critical temperature
%with the size, in both PBC and FBC.

Equations (\ref{eq:bc_finite_size_effect}) and (\ref{eq:tc_finite_size_effect}) 
thus yield the appropriate scaling form for the IF$XY$ model:
In PBC, both ordinary and intrinsic effects add up 
to give the finite-size scaling in the form
\begin{equation}
T_m(L)-T_c =c_i L^{-a}+c_p L^{-1/\nu}, \label{eq:PBC scaling}
\end{equation}
whereas in the case of FBC, the two effects give the convergence behavior
in the opposite direction, leading to the form
\begin{equation}
T_m(L)-T_c =c_i L^{-a}-c_f L^{-1/\nu}. \label{eq:FBC scaling}
\end{equation}
Indeed using the value $\nu =0.9$,
% in Ref.~\cite{Granato},
we can fit the peak position data {\em both in PBC and in FBC}
to Eqs.~(\ref{eq:PBC scaling}) and~(\ref{eq:FBC scaling}) 
with $T_c =0$, respectively, to obtain the {\em same} value
\begin{equation}
a=0.28\pm 0.05.
\end{equation}
The resulting scaling curves are displayed in Fig.~\ref{fig:peak}(b), 
giving strong evidence for $T_c =0$.

\begin{figure}[H]
\centerline{\epsfxsize=8cm \epsfysize=6cm  \epsfbox{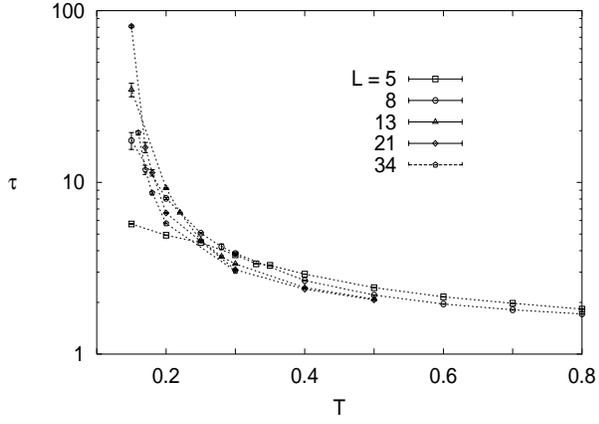}}
\caption[Relaxation time $\tau$ versus the temperature $T$]
{Relaxation time $\tau$ versus the temperature $T$ in the semi-log 
scale for system size $L= 5, 8, 13, 21,$ and $34$. 
Error bars have been estimated from the standard deviation and those data
 points without them have errors not larger than the size of
 the symbols. The dotted lines are merely guides to the eye.}
\label{fig:relaxation_time}
\end{figure}

To confirm this, we also study the relaxation dynamics of
an $L\times L$ RSJ array under PBC in both directions~\cite{BJKim}.
To obtain dynamical behavior, we integrate directly the equations of motion
with the time step $\Delta t =0.05$, starting from
random initial configurations and taking averages over $300$ to $30000$
independent runs.
In the limit $t\rightarrow \infty$, the energy function is expected to approach
the equilibrium value at given temperature $T$, allowing us to estimate
the relaxation time $\tau$ of the energy function.
The obtained behavior of $\tau$ is shown in Fig.~\ref{fig:relaxation_time}
for various sizes $L=5, 8, 13, 21,$ and $34$. 
Notice that for $T>0.25$ the data do not depend appreciably on the system size,
which explains why the size effects were not observed in Ref.~\cite{Granato}. 
Still the scaling of $IV$ characteristics, not relying much on the system size 
in the temperature region probed, can give an accurate value
of the exponent $\nu$~\cite{Granato}.

To examine the finite-size effects in the relaxation time data,
we construct the suitable scaling form of the relaxation time.
%Unlike the ordinary case, where the dynamic exponent $z$ relates 
%the relaxation time to the correlation length via $\tau\sim\xi^z$,
In the presumed zero-temperature glass transition, 
the relaxation has an activated form, and the relaxation time
diverges exponentially as the temperature approaches zero, 
taking the form~\cite{DSFisher91,Hyman}
\begin{equation}
\tau = A \exp(\Delta E/T).
\end{equation}
Here $\Delta E$ is the typical barrier that a vortex should overcome to
move across 
%the distance of the order of 
the correlation length $\xi$, 
and scales as $\Delta E\sim \xi^\psi$ with the barrier exponent $\psi$.
%%%%
%In the ordinary case of a zero-temperature transition, 
%the correlation length is expected to display the divergence
%\begin{equation}
%\xi\sim T^{-\nu}.
%\end{equation}
%
In the presence of intrinsic finite-size effects, 
%on the other hand, 
%the system of finite size exhibits nonzero transition temperature $T_c(L)$,
%which becomes lower with the system size $L$, although the transition is 
%of course rounded due to ordinary finite-size effects.
%Accordingly, 
which lead to the size-dependent transition temperature $T_c(L)$,
the correlation length should display the scaling behavior
\begin{equation}
\xi = L f(L^{1/\nu}[T-T_c(L)])
\end{equation}
with the appropriate scaling function $f(x)$ such that
$f(x\rightarrow 0) \rightarrow \rm{constant}$ and
$f(x\rightarrow \infty) \rightarrow x^{-\nu}$.
%which leads to the scaling of the relaxation time $\tau$ in the form
%\begin{equation}
%\log \tau \sim \Delta E(T)\sim [T-T_c(L)]^{-\psi \nu}.
%\end{equation}
%Since the system size is finite here, $\tau$ does not actually diverge at $T=T_c$ 
%but shows the ordinary finite-size effects.
The corresponding finite-size scaling of the relaxation time thus
takes the form
\begin{equation}
\log \tau = L^\psi g(L^{1/\nu} [T-T_c(L)])
\label{eq:tau_finite_size_scaling}
\end{equation}
with the scaling function $g(x) \propto [f(x)]^{\psi}$.

\begin{figure}
\centerline{\epsfxsize=8cm \epsfysize=6cm  \epsfbox{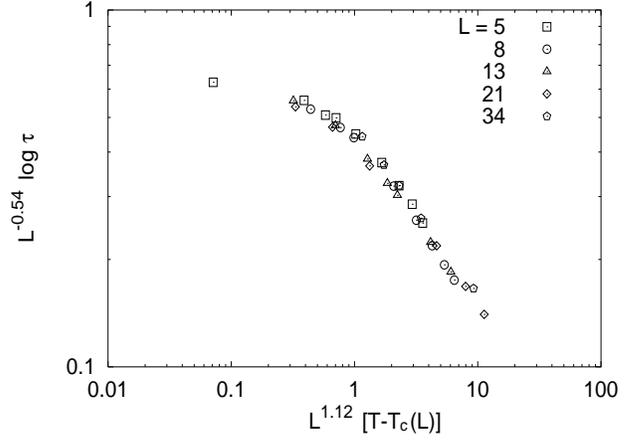}}
\vspace{0.3cm}
\caption [Scaling plot of the relaxation time of the irrationally frustrated 
XY model] 
{Fit of the relaxation time data to the finite-size-scaling 
formula given by Eq.~(\ref{eq:tau_finite_size_scaling}), with
$\psi=0.54$ and $1/\nu=1.12$.  The critical temperature $T_c(L)$ has
been obtained from the specific heat data.
}
\label{fig:tau_scaling}
\end{figure}

Figure~\ref{fig:tau_scaling} presents the scaling plot of the relaxation 
time $\tau$ of the IF$XY$ model, showing that the data indeed 
fit well to the finite 
size-scaling form in Eq.~(\ref{eq:tau_finite_size_scaling}). 
The critical temperature $T_c(L)$ at each size $L$ has been estimated from
the specific heat data.
%, with the contribution of the intrinsic
%finite-size scaling effects given by $c_1 L^{-a}$ 
%in Eq.~(\ref{eq:PBC scaling})
%taken into account.
>From this scaling with $\psi = 0.54$, 
we obtain the exponent $\nu= 0.89\pm 0.2$, 
which is in good agreement with the results of the specific heat data 
and of the $IV$ characteristics in Ref.~\cite{Granato}.

%%%%%
%Finally, we discuss the difference between these intrinsic finite-size effects 
%and the finite-size effects of a weakly first-order transition.
%In the case of a weakly first-order transition, the second-order type of 
%scaling can be observed whenever the correlation length of the system is of the
%order of the system size $L$~\cite{Privman,Peczak}. 
%At a given temperature the correlation length of the system is fixed, 
%and the second-order type of scaling will disappear at the size exceeding 
%the correlation length.
%In the case of intrinsic finite-size effects, on the other hand, 
%as the system size is increased, the approximate rational frustration
%determining the transition is replaced by the approximants more close to 
%the actual irrational, to satisfy Eq.~(\ref{eq:size_criterion}).
%Accordingly, the correlation length does not remain fixed as the size is
%increased, and the scaling behavior is expected to persist, in agreement with
%the obtained data and thus supportive of 
%the intrinsic finite-size effects in the IF$XY$ model.
%
In summary, we have studied the finite-size scaling of the specific
heat and the relaxation time in the two-dimensional irrationally
frustrated $XY$ model.  
This has revealed intrinsic finite-size effects and 
a zero-temperature glass transition in the thermodynamic limit,
thus resolving the controversy as to the existence of
a finite-temperature transition.
%Unlike the latter, 
%the intrinsic effects tend to lower the transition temperature 
%with the system size, regardless of the boundary conditions.
%This in turn suggests a zero-temperature glass transition in the
%thermodynamic limit, resolving the controversy . 
Since the analytic argument for the intrinsic effects 
is rather general~\cite{MYChoi87}, we believe the conclusion to
be valid for any irrationals; this is indeed supported by the preliminary 
results for the silver mean case, 
the detailed investigation of which is left for further study.
Finally, we point out that for generic values of frustration, a 
quasiperiodic array displays behavior similar to that of a 
periodic array with irrational frustration~\cite{Chung}.  
This suggests the presence of intrinsic finite-size effects 
also in the quasiperiodic array, as manifested by the apparent glass behavior reported in Ref.~\cite{quasi}. 

MYC thanks D.J. Thouless for the hospitality during his stay at University of 
Washington, where part of this work was accomplished.  
SYP thanks D.R. Nelson, D.S. Fisher, and S. Teitel for useful discussions
and acknowledges the fellowship from the Rotary Foundation. 
This work was supported 
in part by the NSF Grant DMR-9815932, by the Ministry of Education through 
the BK21 Program (MYC), and by the KRF Grant 99-005-D00034 (JSC).

%Figures

\end{multicols}


\begin{references}
\bibitem{footnote} Similar behavior was also reported in the approach 
using the approximate rational frustration given by the first few of 
the Fibonacci sequence. See Ref.~\cite{Denniston}.
\end{references}
\end{document}